\documentclass[twocolumn]{article}
\usepackage[utf8]{inputenc}

\usepackage{hyperref}
\usepackage{amsmath}

\newcommand{\eq}{\begin{equation}}
\newcommand{\eqx}{\end{equation}}
\newcommand{\eqn}{\begin{eqnarray}}
\newcommand{\eqnx}{\end{eqnarray}}
\newcommand{\f}[2]{\frac{#1}{#2}}

\newcommand{\cor}[1]{\left\langle{#1}\right\rangle}

\newcommand{\sg}{\sigma}

\newcommand{\Dl}{\Delta}

\newcommand{\bt}{\beta}

\title{Entropy from Machine Learning}

\author{Romuald A. Janik\thanks{e-mail: {\tt romuald.janik@gmail.com}} \\ \\ 
\small 
Jagiellonian University,\\\small
Institute of Physics\\\small
ul. {\L}ojasiewicza 11,\\ \small
30-348 Krak{\'o}w,\\\small
Poland}
\date{}

\usepackage{fullpage}
\usepackage{graphicx}

\begin{document}

\maketitle

\begin{abstract}
We translate the problem of calculating the entropy of
a set of binary configurations/signals into a sequence of supervised classification tasks. Subsequently, one can use virtually any machine learning classification algorithm for computing entropy. This procedure can be used to compute entropy, and consequently the free energy directly from a set of Monte Carlo configurations at a given temperature. As a test of the proposed method, using an \emph{off-the-shelf} machine learning classifier we reproduce the entropy and free energy of the 2D Ising model from Monte Carlo configurations at various temperatures throughout its phase diagram. Other potential applications include computing the entropy of spiking neurons or any other multidimensional binary signals.
\end{abstract}

\section{Introduction}

The problem of estimating entropy of high dimensional binary configurations or signals is ubiquitous in many disciplines.
In physics, we very often have at our disposal a set of configurations of some physical system generated by a Monte Carlo simulation
at a given temperature $T_0$. This data is very much geared towards computing   expectation values of various operators or their correlation functions, however obtaining
the entropy or free energy of the system is far from trivial. Indeed, to the best of our knowledge, there is no known way to compute the entropy directly from
these configurations~\cite{COMMENT} even for a system of a quite moderate size (e.g. for a $20 \times 20$ lattice).
The goal of the present paper is to propose machine learning based methods
which would allow to make such a computation.

Due to the lack of a direct method to obtain entropy, various indirect approaches have been proposed.

One standard way requires to perform separate Monte Carlo simulations at a series of temperatures from close to $T=0$
up to $T_0$, evaluate the heat capacity $C(T)$ from the variance of
the energy $\sg^2_E(T)$
\eq
C(T) = \f{\partial\! \cor{E}}{\partial T} = \f{\sg^2_E(T)}{T^2} 
\eqx
and obtain the entropy by a numerical integration over~$T$:
\eq
S(T_0) = \int_0^{T_0} C(T) \f{dT}{T}
\eqx
Clearly this process is time consuming and requires performing multiple auxiliary Monte Carlo simulations even if one is interested only in
the entropy of the system at a specific temperature $T_0$.

Another approach is to use Wang-Landau sampling~\cite{WangLandau} for the given hamiltonian to obtain the density of states $g(E)$, and subsequently compute the entropy.
This way does not use the original Monte Carlo configurations at all and requires to perform a quite separate (and conceptually different) Monte Carlo entropic sampling
to evaluate the entropy. 
Needless to say, Wang-Landau sampling requires for the system to be described by an explicit hamiltonian,
which is of course clearly given in the physics context, but requires separate modeling in other contexts (e.g. for neuron spike recordings).
In the latter case, in order to use any of the above approaches, one would have to assume a specific form of an effective hamiltonian describing the system, which is in fact an additional prior input. Determining its free coefficients is an involved problem by itself, even before
trying to evaluate the entropy.  

Indeed, within the field of neuroscience, one is often interested in the dynamics of spiking time series of populations of neurons and in particular in assessing their information content through an estimation of entropy. For small populations,
one can clearly evaluate the entropy directly (see section~\ref{s.entropy} below), however as the number of available neurons grows, the space of allowed
configurations increases exponentially. 
A common approach, as indicated above, is to model the system through an effective
Ising model~\cite{IsingNeurons} (equivalently a Hopfield network~\cite{Hopfield})
\eq
\label{e.maxentising}
H = \sum_{i<j} J_{ij} s_i s_j + \sum_i h_i s_i
\eqx
which provides a maximal entropy description subject to the constraint of reproducing all 1-point and 2-point correlation functions.
Once one solves this ``inverse Ising'' problem of fitting the parameters $J_{ij}$ and $h_i$ \cite{InverseIsing}, one can then use a physics based approach
(like mean field or one of the more precise methods like temperature integration or the Wang-Landau sampling mentioned above) to evaluate the entropy. As emphasized previously, this is however not a
trivial problem even if the exact coefficients $J_{ij}$ and $h_i$ are known.

The aim of this work is to propose a viable method which computes the entropy directly from the original configurations (e.g. the
Monte Carlo configurations at a given temperature, or neuron spike trains, or any other relatively high dimensional binary signal).
In particular, we would like the method to be applicable when the dimensionality of the system is such that there are basically no
repeated configurations. This is in fact the generic situation in physics simulations, where e.g. the Ising model on even a small $20 \times 20$
lattice has of the order of $10^{120}$ possible configurations\footnote{For the $20000$ Monte Carlo configurations at each temperature for the $20 \times 20$ Ising model used in this paper, we checked that indeed all configurations are distinct for temperatures $T\geq 2.7$.}.

\section{Entropy and its estimation}
\label{s.entropy}

Suppose that an $N$-dimensional binary signal $\{x_i\}_{i=1..N}$ comes from a probability distribution
\eq
p(x_1,x_2,\ldots, x_N)
\eqx
Then the (Shannon) entropy\footnote{In this paper we will always use base 2 logarithms, thus the entropy is measured in \emph{bits}.} is given by
\eq
\label{e.shannon}
S = - \!\!\!\!\!\! \sum_{\{x_i\}_{i=1..N}}\!\!\!\! p(x_1,x_2,\ldots, x_N) \log_2 p(x_1,x_2,\ldots, x_N)
\eqx
where the sum is over the $2^N$ allowed configurations. 
In the physics context, the Shannon entropy is of course equivalent to the thermodynamic entropy, when
the probability distribution is given by a Boltmann distribution associated to some hamiltonian:
\eq
\label{e.boltzmann}
p_{Boltzmann}(x_1,x_2,\ldots, x_N) = \f{1}{Z} e^{-\f{1}{T} H(x_1,x_2,\ldots, x_N) }
\eqx
where $T$ is the temperature.

The difficulty in evaluating the entropy directly from the definition~(\ref{e.shannon}) has two sources. Firstly, even if the explicit functional form (\ref{e.boltzmann})
is known and the hamiltonian $H(x_1,x_2,\ldots, x_N)$ is easy to evaluate as is usually the case,
the computation of the normalizing constant $Z$ (i.e. the partition function in physics) and the summation
over configurations in (\ref{e.shannon}) is intractable by brute force. Hence the need for the theoretically much more involved approaches
mentioned in the introduction.

Secondly, especially outside the physics context, we do not have at our disposal the functional form (\ref{e.boltzmann}) with a simple \emph{known} hamiltonian, and the problem lies in reliably estimating
the probability distribution $p(x_1,x_2,\ldots, x_N)$ given a set of $n$ samples from the distribution.
For relatively small $N$ and a sufficient number of samples, one can directly count the number of times $k$
a given configuration $\{x_i\}_{i=1..N}$ occurred and set
\eq
\label{e.plugin}
p(x_1,x_2,\ldots, x_N) = \f{k}{n}
\eqx
Substituting this back into (\ref{e.shannon}) gives the plug-in (or Maximal Likelihood) estimator.
Taking into account the bias for small sample sizes/occurrence counts leads to a variety
of improved estimators like the Miller-Madow \cite{MillerMadow}, Chao-Shen \cite{ChaoShen}, Grassberger \cite{Grassberger}, James-Stein \cite{Hausser}
and others. Further refinements involve Bayesian estimators based on Dirichlet priors (see a summary and references in \cite{Hausser}),
and the most refined estimators obtained by fitting a mixture of Dirichlet distributions:
the Nemenman-Shafee-Bialek (NSB) estimator \cite{NSB} and the CDM estimator of \cite{CDM}.

Another quite different approach is based on an expansion of the Shannon entropy in terms of mutual information.
This comes from rewriting exactly the entropy in terms of entropies of individual subsystems
and then adding corrections due to first pairwise mutual information and subsequently their higher order
generalizations:
\eq
\label{e.mie}
S = \sum_{i=1}^N S(i) -\sum_{i<j} I_2(i,j) +\sum_{i<j<k} I_3(i,j,k)+\ldots
\eqx
where $S(i)$ is the entropy of the subsystem made up of neuron/spin $x_i$,
$I_2(i,j)$ is the mutual information between the subsystems $x_i$ and $x_j$:
\eq
I_2(i,j) = S(i) + S(j) - S(i,j)
\eqx
etc. Unfortunately, truncation of the expansion (\ref{e.mie}) at say $2^{nd}$ order
may yield huge errors -- even leading to a negative entropy estimate for the Ising model in the intermediate temperature regime. 
In principle this should be improved by higher order terms, however evaluating
them explicitly becomes combinatorially intractable for larger $N$.

Another interesting proposed approach (with somewhat wider goals) \cite{MaxentropyNature}, aims at maximizing the possible entropy 
satisfying a set of information theoretic constraints. Unfortunately this method
also seems to be limited to moderate dimensionalities $N$.  

\section{The proposed method}
\label{s.method}

As indicated in the introduction, we would like our method of estimating entropy to work also in the case of dimensionalities $N$, where
the given samples involve basically only distinct configurations, thus in formula~(\ref{e.plugin}), $k=0$ or $k=1$
and we cannot use any of the subsequent refinements.

Before we present the details, let us comment on the general philosophy of the proposed method.

The methods of entropy estimation based on occurrence counts like~(\ref{e.plugin}) and its variations treat each distinct configuration as a structureless ``atomic'' object. All details of its constituents are thrown away. Our method, in contrast, strives to analyze the internal structure of the configurations and thus can work in the regime where all occurring configurations are distinct. In order to extract the possibly quite complex internal relationships between the constituents, we will use machine learning methods, which, as we will show, appear here in a very natural way.

Our starting point is the exact rewriting of a multidimensional probability distribution in terms of
a product of conditional probabilities
\eq
\label{e.condprobfact}
p(x_1,x_2,\ldots, x_N) = p(x_1) p(x_2|x_1) p(x_3|x_1,x_2) \cdot \ldots
\eqx
Let us first evaluate Shannon's entropy using the above decomposition.
We clearly get a sum of $N$ terms:
\eq
\label{e.decomposition}
S = S_1 + \Delta S_2 + \Delta S_3 + \ldots \Delta S_N
\eqx
The first term is just the entropy of the first neuron/spin:
\eq
S_1 = -p_1 \log_2 p_1 - (1-p_1) \log_2 (1-p_1)
\eqx
where $p_1 \equiv p(x_1=1)$.

The second term is more interesting. We need to evaluate
\eq
\Delta S_2 = -\!\!\!\sum_{\{x_1,x_2\}} \!\! p(x_1, x_2) \log_2 p(x_2|x_1)
\eqx
Since $x_2$ is a binary variable, the conditional probability is completely specified
by the simple function
\eq
p_2(x_1) \equiv p(x_2=1 | x_1)
\eqx
Then an estimate of $\Delta S_2$ based on the dataset $\{x_1^{(k)}, x_2^{(k)}\}_{k=1 \ldots n}$ is
\eq
\label{e.deltastwo}
-\f{1}{n} \sum_{k=1}^n x_2^{(k)} \log_2 p_2(x_1^{(k)}) + (1-x_2^{(k)}) \log_2 (1- p_2(x_1^{(k)}))
\eqx
We observe, that this is exactly the cross-entropy loss in a standard supervised classification problem
where we treat the value of the neuron/spin $x_2$ as a class label $y$ (which can be either 0 or 1),
and we try to predict its probability based on the value of neuron/spin $x_1$.
Similarly for $\Delta S_3$ we get
\eqn
-\f{1}{n} \sum_{k=1}^n &&\hspace{-0.5cm} x_3^{(k)} \log_2 p_3(x_1^{(k)}, x_2^{(k)}) + \nonumber\\
 &&\hspace{-0.8cm} + \left(1-x_3^{(k)}\right) \log_2 \left(1- p_3(x_1^{(k)}, x_2^{(k)})\right)
\label{e.deltasthree}
\eqnx
where now we need to predict the probability of $x_3=1$ in terms of
the values of $x_1$ and $x_2$.

Thus the decomposition (\ref{e.decomposition}) corresponds exactly to a sum of cross-entropy losses
of a sequence of iterative supervised classification problems where we predict the probability of
$x_j=1$ given the values of the previous\footnote{According to the given fixed ordering $x_1, x_2, \ldots, x_N$.} 
spins $x_1, x_2, \ldots, x_{j-1}$. 

Note, however, that these auxiliary classification problems have a somewhat different flavour from the typical ones encountered in machine learning.
There, we expect usually the target class to be completely determined by the features (predictors), like whether a given image represents a cat or a dog. So the ideal classifier should attain zero cross-entropy loss.
In our case, we expect generically for the best possible classifier to achieve only some positive cross-entropy loss.
It is exactly this nonzero cross-entropy which summarizes the contribution of the $j$-th spin/neuron to the total entropy.

Indeed, suppose that spin $x_j$ is completely independent of $x_1, x_2, \ldots, x_{j-1}$.
Then the conditional probability 
\eq
p_j(x_1, x_2, \ldots, x_{j-1}) \equiv p(x_j=1|x_1, x_2, \ldots, x_{j-1})
\eqx
will just be a constant equal to the marginal $p(x_j=1)$ and $\Delta S_j$ will
be equal to the entropy of the neuron/spin $x_j$ by itself.

If, on the other hand, $x_j$ would be completely determined by $x_1, x_2, \ldots, x_{j-1}$,
we would expect the ideal classifier to give vanishing cross-entropy and $\Delta S_j=0$.

Let us emphasize that the above procedure of computing the entropy depends explicitly on the predefined ordering of the particular 
neurons/spins $x_1,x_2,\ldots, x_N$, as the auxiliary classification problems for two orderings are completely different.
On the other hand, the final outcome, the entropy of the system,
is clearly independent of any ordering. This property may be a nontrivial cross-check of the 
quality of the employed machine learning classifier, hence it would be beneficial to evaluate
the entropy for a couple of permutations of the neurons/spins.

Let us note that in the limit of infinite data, the entropy estimate~(\ref{e.decomposition}) using some concrete classification algorithm will approximate the true entropy from above. This can be easily seen e.g. for the difference $\Dl S_2^{true} - \Dl S_2^{classifier}$ equal in the infinite data limit
\eq
\!\!\!\sum_{\{x_1,x_2\}} \!\! p(x_1, x_2) \log_2 \f{p_{classifier}(x_2|x_1)}{p(x_2|x_1)}
\eqx
Using Jensen's inequality this is smaller or equal than
\eq
\log_2 \!\!\!\sum_{\{x_1,x_2\}} \!\! p(x_1, x_2) \f{p_{classifier}(x_2|x_1)}{p(x_2|x_1)}
\eqx
which in turn vanishes due to
\eq
\log_2 \!\!\!\sum_{\{x_1,x_2\}} \!\! p(x_1) p_{classifier}(x_2|x_1) = \log_2 1 = 0
\eqx
A similar argument holds for other terms.
Hence if various classifiers yield different answers, we should chose the one which predicts the
lowest entropy. Of course, for a finite number of samples these theoretical conclusions are not guaranteed to hold.

Finally, as emphasized in more detail at the end of the following section, we should take care how we evaluate the predictions of the fitted classifiers and never evaluate them on the data used for training.

\section{Machine learning considerations}
\label{s.ml}

In section \ref{s.method}, we reinterpreted the decomposition (\ref{e.decomposition}) for
the formula for the entropy (\ref{e.shannon}) in terms of cross-entropy losses of
a sequence of supervised classification problems. This is very general and comes from
the fact that a discriminative supervised classification problem is just
by definition a certain model of the conditional probability distribution
\eq
\label{e.condprob}
p(y|x_1,x_2,\ldots,x_j)
\eqx
where $y$ is a binary (target/class) variable and $x_1,x_2,\ldots,x_j$ are some predictors/features.

The reason, why this interpretation is fruitful is that we would like to estimate (\ref{e.condprob})
in a data driven way based on the given set of binary signals (representing e.g. Monte-Carlo spin configurations,
neuron spiking time series etc.) and the field of Machine Learning provides in fact a very wide variety
of algorithms which are specifically designed to model \emph{very complex} conditional probability distributions\footnote{A classical example would be the probability that an image shows a cat conditioned on the set of pixels of the image.},
and thus can be directly used now to estimate entropy.
The most notable examples are logistic regression, $k$-nearest neighbours, deep neural networks,  random forests, gradient boosted trees,  and others.

Note that the relevance of the machine learning algorithm for computing entropy is indicated not by classification accuracy, but rather by a good estimate of probability. Thus e.g. Naive Bayes is not expected to be a good choice, Support Vector Machines (SVM) by default do not provide probabilities at all  and deep neural networks may require some caution (cf.~\cite{Khan}).

The freedom in the choice of classifier means that we have at our disposal a whole range of methods for computing entropy. This variety may be very useful when we have some prior knowledge about the given binary signals. In the case when the number of samples is not so large, Bayesian models/inference may be very effective. One can also use any of the techniques
developed within machine learning for feature selection as an ingredient for the
procedure of entropy estimation.

In this paper we will not attempt to investigate any of such refinements, but rather, as a proof of concept of the method, use a standard \emph{off-the-shelf} classifier and use it with basically default settings

A very versatile nonlinear classifier which will be used in the numerical experiments
in the present paper is the so-called \emph{gradient boosted tree} classifier. 
Specifically, we chose {\tt xgboost} -- its very efficient\footnote{For computational speed, in this paper we always use the settings {\tt tree\_method='hist'} and {\tt n\_jobs=-1}.} variant and implementation \cite{xgboost}.
In some cases, we will also use logistic regression for comparison.

Before we close this section, we need to address a remaining important issue
concerning the evaluation of the predictions $p_j(x_1,x_2,\ldots,x_{j-1})$ used in formulas like
(\ref{e.deltastwo}) or (\ref{e.deltasthree}). In Machine Learning one has to avoid ``overfitting'',
where the classifier will fit random noise in the data. This is especially dangerous for the more complex
nonlinear classifiers, which are incidentally of the most interest in the present context. Moreover, predicting directly on the training data used to fit the classifier would strongly skew the predicted probabilities and thus would tend to underestimate the entropy if we have a finite amount of data
(as is always the case).

A standard way to mitigate this problem is to partition the dataset into $k$ (at least $k=2$ but usually $k=5$) parts (folds),
train the classifier on the sum of $k-1$ parts, but compute predicted probabilities only on the remaining  unseen $k$-th fold.
Then repeat the process $k-1$ times, holding out another fold, until one gets predictions for all datapoints.
Then the cross-entropy loss like (\ref{e.deltastwo}) or (\ref{e.deltasthree}) entering (\ref{e.decomposition}) should be computed only using these held-out predictions\footnote{In order not to be biased by a specific choice of the partition of the dataset into $k$ folds, we pick a different shuffled partition for each auxiliary classification problem.}.

\section{The importance of nonlinearity}

In this section, we will consider synthetic datasets where some binary features are
functionally dependent on others. 

Let $X_1$ be given by 50 independent random binary variables, each with probability $0.5$, and $X_2$ be defined similarly. We will now form four variants of a third set
of binary variables $X_3$ which are given respectively by $\text{NOT}\ X_1$, $X_1\ \text{OR}\ X_2$, $X_1\ \text{AND}\ X_2$ and $X_1\ \text{XOR}\ X_2$. 
The final datasets will be obtained by sampling the
concatenation of $X_1$, $X_2$ and one of the variants of $X_3$. 
Hence in total
we have 150 binary variables, whose entropy is clearly equal to $100.0$ bits irrespective of the 
choice of $X_3$.
We will now perform a random reordering of the variables, take $10000$ samples and apply
our proposed method for computing the entropy.

It is illuminating to look at the answers obtained by using two standard classifiers:
logistic regression\footnote{We use the implementation from {\tt scikit-learn} with default parameters.} and gradient boosted trees ({\tt xgboost}). The former is essentially a baseline linear classifier, while the latter is, as mentioned earlier, a fully nonlinear one. The results are shown in Table~\ref{tab.dependence}.

\begin{table}[h]
    \centering
    \begin{tabular}{|c|c|c|c|c|}
    \hline
         &  NOT & OR & AND & XOR \\
    \hline
    lr    & 100.68 & 101.23 & 101.26 & 151.05 \\
xgb(100)  & 100.74 & 100.74 & 100.71 & 122.27 \\
xgb(200)  & $\cdot$& $\cdot$& $\cdot$& 112.60 \\
xgb(400)  & $\cdot$& $\cdot$& $\cdot$& 105.79 \\
\hline
    \end{tabular}
    \caption{The results for estimating the entropy for the four datasets, with logistic regression,
    {\tt xgboost} with the default number of trees (100), and with 200 and 400 trees respectively. The exact answer for all datasets is $100.0$ bits.}
    \label{tab.dependence}
\end{table}{}

Unsurprisingly, the interdependence structure of the XOR dataset cannot be captured by a linear model, and logistic regression indeed very strongly overestimates the entropy. The nonlinear {\tt xgboost} classifier is definitely better, but the structure of the dataset requires increasing the complexity of the model (by increasing the number of trees from the default 100 to 400 and possibly more\footnote{Note, however, that increasing the complexity of the model is not always beneficial. See e.g. the case of the 2D Ising model in the following section.}) in order to correctly estimate the entropy.

This example serves to illustrate that the flexibility of our proposed method in the choice of the specific machine learning classifier can yield interesting insights into the data. By comparing different classifiers we can e.g. assess the complexity of nonlinear interdependence of our configurations or signals. Alternatively, we can sometimes restrict ourselves to a simpler and faster model if the more complex model does not yield a significantly lower entropy estimate.

\section{Ising model entropy and free energy from Monte Carlo configurations}

As a nontrivial cross check of the proposed method we will evaluate
the entropy of the 2D Ising model on a $20 \times 20$ periodic lattice
directly from sets of 20000 configurations obtained using Monte Carlo sampling
for temperatures ranging from $T=1.0$ to $T=4.0$, thus spanning
both the low and high temperature phase of the infinite volume theory and the critical phase transition in between.

The 2D Ising model is especially interesting as a testing ground
for the evaluation of entropy for a number of reasons. Firstly,
there are analytical exact formulas for the entropy and free energy
of the model put on any $L \times L$ lattice (see the formulas in Appendix~A). Secondly, it is not a trivial system but exhibits
a phase structure with two distinct phases separated by
a second order phase transition. Thirdly, it has a huge number of possible configurations $2^{L^2}$, thus making it a challenging testing ground for entropy estimation. Fourthly, it is a quintessential example for Monte Carlo simulations, thus the
direct computation of the entropy and free energy just from the Monte Carlo configurations would be an interesting proof of concept of the proposed method. Finally, generalizations of the Ising model with arbitrary pairwise couplings and inhomogeneous magnetic field (\ref{e.maxentising})
are commonly employed as maximal entropy models of neuron spiking in neuroscience.

\begin{table}[t]
    \centering
    \begin{tabular}{|c|c|c|r|}
    \hline
      {\tt max\_depth}   &  {\tt n\_estimators} & $S(T=T_c)$ & time \\
      \hline
       3  & 25 &   0.44125 & 7.7\\
       3  & 50 &   0.42815 & 12.1\\
       3  & 75 &   0.42868 & 16.5\\
       3  & 100 &  0.42960 & 21.0\\
       3  & 200 &  0.43330 & 38.7\\
       \hline
       5  & 100 &  0.43885 & 37.2\\
       5  & 200 &  0.45196 & 71.2\\
       \hline
       \multicolumn{2}{|c|}{logistic regression} & 0.43518 & 7.9 \\
       \hline
      \multicolumn{2}{|c|}{\bf exact} & {\bf 0.42468} & \\
      \hline
    \end{tabular}
    \caption{The estimated entropy per spin computed from 20000 Monte Carlo samples at $T=T_c$ for various choices of hyperparameters of {\tt xgboost} as well as logistic regression for completeness. The running time is in minutes on a 6-core desktop. The exact answer is given for a $20\times 20$ lattice (see Appendix~A).}
    \label{tab.hyper}
\end{table}{}

Since the 2D Ising model has a clear spatial structure, and we would like to test our method in a more general context, we will randomly permute the $400$ spins to set the ordering for (\ref{e.condprobfact}).

Before we present the predictions for the whole range of temperatures it is instructive to look at the dependence of the outcome on various factors.

Firstly, two key hyperparameters of the gradient boosted trees algorithm are the number of trees ({\tt n\_estimators}) and their maximal depth 
({\tt max\_depth}). In Table~\ref{tab.hyper}, we show the dependence of the entropy estimate on these choices at the
critical point $T=T_c \sim 2.2691853$.

We see that for 25 trees, the model is too simple, while for
increasing depth and number of trees it starts overfitting. Thus, for all the remaining experiments we choose the default {\tt max\_depth=3} and just decrease the number of trees from the default $100$ to~$50$. This is also clearly better than a logistic regression baseline.

It is also interesting to analyze the dependence of the entropy estimate on the number of Monte Carlo samples. The results for 
$n=10000$, $n=20000$ and $n=40000$ are shown in Table~\ref{tab.n}.
The results get better as we gain access to more data.
This motivated us to compute a linear extrapolation in $1/n$, although there are
no known theoretical grounds for a specific functional dependence.
This is, in fact, an interesting problem for further study.

\begin{table}[h]
    \centering
    \begin{tabular}{|c|c|}
    \hline
        $n$ & $S(T=T_c)$ \\
        \hline
      10000   &  0.42989\\
      20000   &  0.42815\\
      40000   &  0.42748\\
      \hline
      extrapolated & 0.42661\\
      \hline
      {\bf exact} & {\bf 0.42468}\\
      \hline
    \end{tabular}
    \caption{The estimated entropy per spin at $T=T_c$ as a function of the number of Monte Carlo samples, together with a linear extrapolation in $1/n$.}
    \label{tab.n}
\end{table}

As emphasized in section \ref{s.ml}, it is important to estimate the probabilities by predicting on an unseen test set, especially for more complex classifiers, so we always use $k$-fold cross validation.
Increasing $k$, increases the amount of training data for the machine learning algorithm on each fold, thus leading to better results. The drawback, however, is that the running time increases as one has to run each classifier $k$ times to obtain predictions for all samples. In Table~\ref{tab.cv} we show this dependence on~$k$. In all other simulations we take $k=5$.

\begin{table}[h]
    \centering
    \begin{tabular}{|c|c|c|}
    \hline
        $k$ & $S(T=T_c)$ & time\\
        \hline
        2 & 0.42935  & 2.7\\
        3 & 0.42857  & 5.2\\
        4 & 0.42829  & 7.4\\
        5 & 0.42815  & 9.1\\
      \hline
      {\bf exact} & {\bf 0.42468} &\\
      \hline
    \end{tabular}
    \caption{The estimated entropy per spin computed from 20000 Monte Carlo samples at $T=T_c$ for various number of cross validation folds together with the running time in minutes.}
    \label{tab.cv}
\end{table}

As emphasized in section~\ref{s.method}, the procedure for estimating the entropy depends explicitly on the ordering of the variables (spins/neurons) as the auxiliary classification problems (i.e. which spin/neuron to predict based on which subset of spins/neurons) are quite different for different orderings. Yet the sum of the cross entropy losses for all the classification problems should not depend on the ordering. This is a nontrivial consistency check of the method and may serve to assess whether the chosen machine learning classifier is adequate for the given dataset and to estimate the lower bound on the error\footnote{Of course, the difference between the entropy estimate and the true entropy may be much higher than that, as the classifier may \emph{systematically} fail to identify all predictive regularities in the dataset, c.f. logistic regression for the XOR dataset in the previous section.}.

In Table~\ref{tab.order}, we show the results obtained for a natural row-wise ordering, five different random orderings as well as two orderings based on correlation. For the {\it max corr.} one, we iteratively pick a spin which is most correlated\footnote{As measured by the sum of absolute values of correlation coefficients $\sum_{j\in previous} |C_{ij}|$.} with the ones chosen earlier, while for the {\it min corr.} ordering we pick the least correlated spin. We see that the results for various orderings are consistent between themselves.

\begin{table}[t]
    \centering
    \begin{tabular}{|c|c|}
    \hline
    ordering & $S(T=T_c)$ \\
        \hline
    row-wise            &  0.42821\\
    max corr.           &  0.42818\\
    min corr.           &  0.42820\\
    \hline
    random (seed 0)     &  0.42815\\
    random (seed 137)   &  0.42816\\
    random (seed 555)   &  0.42810\\
    random (seed 1621)  &  0.42812\\
    random (seed 4567)  &  0.42817\\
      \hline
      {\bf exact} & {\bf 0.42468} \\
      \hline
    \end{tabular}
    \caption{Entropy estimates computed for various orderings of the spins.}
    \label{tab.order}
\end{table}

\begin{figure}[t]
\hspace{-0.2cm}\includegraphics[width=8.25cm]{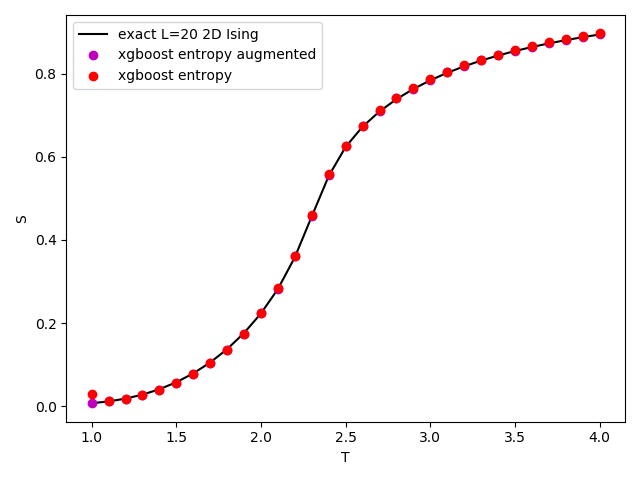}
\caption{Entropy per spin computed using {\tt xgboost} classifier compared with the exact answer. The magenta data point at $T=1.0$ uses data augmentation (see text).} 
\label{fig.entropy}
\end{figure}

With the above exploratory analysis done, we compute the entropy for the range of temperatures $T=1.0$ to $T=4.0$ from 20000 Monte Carlo configurations (at each temperature) and compare with the exact answer for the Ising model on the $20 \times 20$ periodic lattice (see Appendix~A for explicit formulas). 
We use 5-fold cross validation, {\tt max\_depth=3} and {\tt n\_estimators=50}.
We use a random ordering with seed 0.
The results are shown in Fig.~\ref{fig.entropy}. 

\begin{figure}[t]
\hspace{-0.2cm}\includegraphics[width=8.25cm]{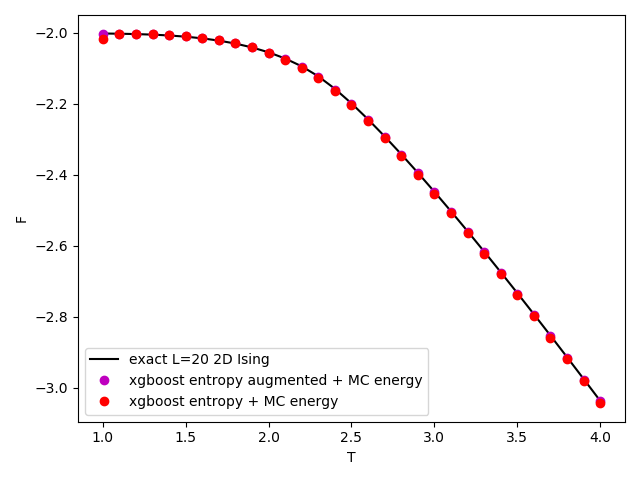}
\caption{Free energy per spin computed using {\tt xgboost} classifier compared with the exact answer. The magenta data point  at $T=1.0$ uses data augmentation (see text).} 
\label{fig.free}
\end{figure}

Once we have the entropy, we can immediately compute the free energy from
\eq
F= \cor{E} -T S
\eqx
where the expectation value of the energy is trivial to compute from the Monte Carlo configurations. The results are shown in Fig.~\ref{fig.free}.

Note that there is a significant deviation for the lowest temperature $T=1.0$. There, the system is in the ordered phase where all spins are predominantly oriented in the same direction. This means that the classification problems are very strongly imbalanced. In fact, it is quite surprising that the {\tt xgboost} classifier works so well for slightly higher temperatures in the ordered phase.

One way to alleviate the problem is to use more data.
Instead of generating further Monte Carlo samples,
one can instead use the very standard machine learning procedure of data augmentation, namely constructing additional samples based on the original data (e.g. cropping and flipping images in image classification tasks). In the context of the Ising model, we can rotate the spin configurations by multiples of $90^{\circ}$ or
perform arbitrary (periodic) translations. Note that from the point of view of the auxiliary classification problems, this yields completely novel additional data. For the magenta points in the figures, we added multiples of $90^{\circ}$ rotations of the original 20000 samples, giving all together 80000 data points. We see that the entropy estimate is now much better.

Of course, data augmentation is only possible if the relevant system has some known symmetries. This is a very common situation for various physical systems but does not occur e.g. for recordings of spiking neurons. In the latter case data augmentation is impossible and one would have to acquire more real data.

\section{Discussion}

In this paper we translated the problem of computing the entropy of a set of binary configurations or signals into performing a sequence of supervised classification tasks, whose sum of cross-entropy losses provides an estimate of the entropy.
We showed that the method is powerful enough to reproduce quite well the entropy and free energy of the 2D Ising model directly from Monte Carlo configurations. 

This framework is very general and allows to use the whole machine learning toolbox with a wide range of diverse machine learning classifier algorithms for computing entropy.
This flexibility may be used, on the one hand, to chose a classifier most suited to the general structure of the data of interest. And, on the other hand, a comparison of the entropy estimates due to different classifiers may lead to interesting insights into the data, like being an indication of the inherent complexity/nonlinearity of the signals.

We hope that these methods will be very useful in physics, where they may be used to estimate the entropy and free energy directly from a set of Monte Carlo configurations.
Within neuroscience, they may be an aid in estimating the entropy of larger populations of spiking neurons as well as other brain signals. Comparing the entropy of a maxentropy model like~(\ref{e.maxentising}) with the entropy obtained from the original signal may be an aid in estimating the applicability of the specific model to the observed data.

There are numerous directions for further study. It should be possible to extend this approach to other information theoretic quantities. 
It would be interesting as well to develop extensions for differential entropy (of continuous probability distributions). Another direction would be to investigate the optimal machine learning algorithms and/or feature selection procedures in various specific contexts like the case of very high dimensionality.

\bigskip

\noindent{}{\bf Acknowledgments.}
I would like to thank Przemysław Witaszczyk for the ongoing collaboration which motivated this investigation and Piotr Białas for interesting discussions on entropy.
This work was done in preparation for the Foundation for Polish Science (FNP) project \emph{Bio-inspired Artificial Neural Networks} POIR.04.04.00-00-14DE/18-00.

\bigskip

\noindent{}{\bf Code and data availability.} All code for reproducing the numerical experiments performed in this paper is available at \href{https://github.com/rmldj/ml-entropy}{\tt github.com/rmldj/ml-entropy}. The 2D Ising Monte Carlo configurations were generated by \href{https://github.com/rmldj/ising}{\tt github.com/rmldj/ising}, a minimally modified fork of \href{https://github.com/zeehio/ising}{\tt github.com/zeehio/ising}. The generated Monte Carlo configurations are also available at \href{https://doi.org/10.5281/zenodo.3457123}{\tt doi:10.5281/zenodo.3457123}. 

\appendix

\section{Exact solution of the 2D Ising model on a $L\times L$ periodic lattice}

Apart from Onsager's exact solution of the 2D Ising model in the thermodynamic limit \cite{Onsager}, there exists also an explicit solution on a finite size $L \times L$ lattice with periodic boundary conditions due to Kaufman~\cite{Kaufman}.
Here we give the formulas as in \cite{appendix}, where the potential branch cut ambiguities in the original formulas have been resolved.

The exact partition function is given by
\eq
Z(L,\bt) = \f{1}{2} \left( 2 \sinh(2\bt) \right)^{\f{L^2}{2}}
\cdot 
\sum_{i=1}^4 Z_i
\eqx
where $\bt=1/T$ and
\eqn
Z_1 \!\!\!&=&\!\!\! 2^L \prod_{r=0}^{L-1} T_{\f{L}{2}}( c_{2r+1} )  \\
Z_2 \!\!\!&=&\!\!\! 2^L \prod_{r=0}^{L-1} U_{\f{L}{2}-1}( c_{2r+1} )
\cdot \prod_{r=0}^{\f{L}{2}-1} \left( c_{2r+1}^2-1 \right)\\
Z_3 \!\!\!&=&\!\!\! 2^L \prod_{r=0}^{L-1} T_{\f{L}{2}}( c_{2r} )  \\
Z_4 \!\!\!&=&\!\!\! 2^L \prod_{r=0}^{L-1} U_{\f{L}{2}-1}( c_{2r} )
\cdot \prod_{r=1}^{\f{L}{2}-1} \left( c_{2r}^2-1 \right) \cdot \\
& & \cdot \left( \cosh^2 (2\bt) - \coth^2 (2\bt) \right)
\eqnx
and
\eq
c_l = \cosh(2\bt) \coth(2\bt) - \cos\f{\pi l}{L}
\eqx
In the above formulas $T_n(.)$ and $U_n(.)$ are Chebyshev polynomials.
The total entropy is then given by
\eq
\label{e.slogz}
S = \log Z - \bt \f{\partial}{\partial \bt} \log Z
\eqx
while the free energy is
\eq
F = -\f{1}{\bt} \log Z
\eqx
In the paper we present results per spin, hence these quantities are divided by $L^2$. Moreover, the entropy is counted in bits, hence (\ref{e.slogz}) is further divided by $\log 2$.

\end{document}